\documentclass[12pt]{article}
\usepackage[pdftex]{graphicx}
\usepackage{amsmath}
\usepackage{url}
\usepackage{amssymb}

\title{\bf Astronomical distances and velocities and special relativity}
\author{Germano D'Abramo\\
{\small Ministero dell'Istruzione, dell'Universt\`a e della Ricerca,}\\
{\small 00041, Albano Laziale, RM, Italy}\\
{\small E--mail: {\tt germano.dabramo@gmail.com}}}

\date{\small {\em Annales de la Fondation Louis de Broglie} {\bf 45}(1) (2020)}

\begin{document}

\maketitle

\begin{abstract}
We show that some primary special relativity effects, which are
believed to be hardly detectable in everyday life, such as time dilation,
relativistic Doppler effect, and length contraction, should tangibly and
spectacularly show up here on the Earth. They should occur 
in ordinary observations of known astronomical phenomena, also when these 
phenomena involve astronomical systems that move with very low velocities
relative to us but are very distant. We shall do that by providing
a reanalysis of the so-called Andromeda paradox and by revisiting the standard 
explanation of the muon lifetime dilation given when this phenomenon is
observed from muon's perspective. 
Ultimately, we shall show that if Lorentz transformations (and basically, 
special relativity) are meant to entail real physical consequences, then the 
observable Universe should appear very differently from what we see 
every clear night.\\

\end{abstract}

\section{Introduction}	
\label{intro}
It is well known that most of the primary special relativity effects,
such as time dilation, relativistic Doppler effect, and length contraction, 
become macroscopically observable only when the velocity $v$ of the
physical system, relative to the observer, approaches the speed of light $c$. 
There is one notable exception, though. According to Purcell's explanation of
magnetic forces, the magnetic  force acting upon a single charge moving
parallel to a neutral current-carrying wire (Lorentz force) is, in fact, a
macroscopic manifestation of the relativistic length contraction of the
distances between the moving conduction electrons in the wire, even though the
velocities involved are always $v\ll c$.
The contraction, which is observable only in the reference frame of
the moving single charge, allegedly causes an unbalance in the charge density
of the wire that results in the attraction (or repulsion) of the moving single
charge. However, the present author has already shown~\cite{prob, pur} that 
this mechanical/dynamical approach to the explanation of magnetic forces is 
problematic and we do not deal with it here.

To the author's knowledge, it is less widely known that special 
relativity effects\footnote{To make the picture clearer from here on
out, with `special relativity effects' we actually mean effects which are the 
mathematical consequence of the application of the Lorentz transformations.} 
should macroscopically 
show up also in other physical systems moving with very low relative
velocities ($v\ll c$), provided that they are placed at huge distances $d$
from the observer (with $d/c^2\gtrsim 1\,$s$^2$/m). Astronomical
objects, with their huge distances and fairly high velocities relative to
the Earth, are thus good candidates to actually observe special relativity
effects.

In the following two sections we describe two examples of relativistic effects
which should allegedly show up in plain observations of astronomical 
objects (very distant and/or very fast) made here on the Earth:
the first example is related to the so-called `Andromeda paradox', while in 
the second one we compare the relativistic explanation of the muon retarded
decay, given when the phenomenon is analyzed from the muon reference frame, 
to what we should see from the Earth when we observe relatively fast 
astronomical objects.

\section{The Andromeda paradox}
\label{sec:1}

The Andromeda paradox, also known with the name of 
Rietdijk--Putnam--Penrose argument \cite{rie,put,pen,wi}, gives a colorful 
demonstration that if special relativity is true, then observers moving at 
different relative velocities (any velocity, also non-relativistic) 
have different sets of events that are {\em present} for them. In particular, 
if two people walk past each other in the street and one of the people was 
walking towards the Andromeda galaxy, then the events in this galaxy that are 
simultaneous with the present time of this observer might be hours or even days
advanced of the events on Andromeda simultaneous with the person walking in the
other direction. 

This argument has been introduced in the past to support the philosophical 
stance known as `four-dimensionalism' (or `block Universe' view), namely 
that an object's persistence through time is like its extension through space 
(for an entertaining and accessible presentation of the philosophical and 
physical theories of Time see, for instance, \cite{cra}).

\subsection{Simple derivation of the paradox}
\label{subsec:11}

The Andromeda paradox can be explained by recurring to the {\em planes 
of simultaneity} in the space-time diagram (Minkowski diagram). Here, 
instead, we make use of the plain Lorentz transformations\\

\begin{center}
\begin{tabular}{rl}
$x'=$ & $\frac{x-vt}{\sqrt{1-\frac{v^2}{c^2}}}$,\\
$y'=$ & $y$, \\
$z'=$ & $z$, \\
$t'=$ & $\frac{t-\frac{vx}{c^2}}{\sqrt{1-\frac{v^2}{c^2}}},$
\end{tabular}
\end{center}
\vspace{0.5cm}

\noindent where the non-primed coordinates $(x,y,z,t)$ refer to the reference 
frame assumed to be at rest and $v$ is the velocity of the primed frame 
relative to the non-primed one along the $x$-axis. 

Consider an observer $A$ here on the Earth who moves towards the Andromeda 
galaxy (relative distance $d$, the direction Earth-Andromeda being along the 
$x$-axis) at a relative velocity $v$ with $v\ll c$.
For the sake of derivation, we shall equivalently consider the Andromeda
galaxy as approaching the observer, and thus the
velocity to be inserted in the Lorentz transformations is $-v$.
Observer $B$ is also on the Earth. He is initially close to the place where $A$ 
starts walking, but he is at rest. It is further assumed that the relative
velocity between the Earth and the Andromeda galaxy is negligible, and thus
the relative velocity of observer $B$ relative to Andromeda is taken as
zero.
According to the Lorentz transformations, if $t_A$ and $t_B$ are the present
instants of time of observer $A$ and observer $B$ respectively (with 
$t_A\simeq t_B$, since $v\ll c$ and the observers' clocks can be considered as
continuously synchronized), then the instant of time on Andromeda simultaneous 
with $t_A$ is

\begin{equation}
t_A'=\frac{t_A+\frac{vd}{c^2}}{\sqrt{1-\frac{v^2}{c^2}}},
\label{eq1}
\end{equation}

\noindent while the instant of time on Andromeda simultaneous with $t_B$ 
$(\simeq t_A)$ can be taken as simply $t_B' = t_B$. 

Since the distance $d$ between the Andromeda galaxy and the Earth is huge, we 
have that $\frac{vd}{c^2}$ can be much greater than unity, even with $v\ll c$, 
and then eq.~(\ref{eq1}) can be approximated to

\begin{equation}
t_A'\simeq t_A+\frac{vd}{c^2}.
\label{eq2}
\end{equation}

This has the paradoxical consequence that although observer $A$ and observer 
$B$ always experience the same `present instant' of time ($t_A\simeq t_B$), 
the events on Andromeda simultaneous with observer $A$ are events 
subsequent (instant of time $t_A'\simeq t_A+\frac{vd}{c^2}$) to the events on 
the same galaxy that are simultaneous with observer $B$ (instant of time 
$t_B'=t_B\simeq t_A$). For instance, it might well happen that, in the 
{\em plane of simultaneity} of observer $A$, a supernova has just exploded 
in some part of the Andromeda galaxy while, in the {\em plane of simultaneity} 
of observer $B$, the same event has not yet happened.

\subsection{Going further}
\label{subsec:12}

In the literature, the extent of the paradox's consequences has been partially 
downplayed by noticing that the observers cannot actually see what is happening 
in Andromeda since it is light-years away, and then the paradox is only 
that they have different ideas of what is happening ``now'' in Andromeda. 

We believe that there is more to it. There is something that can be
in principle physically measured. Suppose that, for the sake of argument, 
both observes can live for millions of years and both decide, starting at time 
$t_A\simeq t_B$, to wait an interval of time equal to $d/c$ and see what 
happens. This interval is the time needed by a light signal emitted in the 
Andromeda galaxy to reach the Earth. Please note that observer $A$ does not keep
moving for the whole interval of time $d/c$: observer $A$ is near observer $B$,
moves a bit, and then comes back near to $B$ almost immediately.

Now, the problem is: what will observer $A$ and observer $B$ see after 
the interval of time $d/c$ has passed? Will they see the same events or not?

What observer $A$ sees after the interval $d/c$ are
the events that were simultaneous with the instant of time $t_A$ of 
observer $A$ exactly $d/c$ years ago and we have just seen that 
these events are surely different from the events that were simultaneous with 
the instant of time $t_B$ $(\simeq t_A$) of observer $B$ exactly $d/c$ years 
ago. All this means that after the same interval of time $d/c$ has passed, 
observer $A$ and observer $B$, {\em who are at rest and close to one
another already for a time nearly equal to $d/c$}, will
{\em actually} see different events while observing the very same galaxy at
the very same time here on the Earth (e.g.~observer $A$ detects the explosion
of a supernova and observer $B$ does not). 

Let us linger over this with the following more direct representation.  
Observer Bob is at rest on the Earth, sitting on a bench and staring at the 
Andromeda galaxy (which is $d$ away from the Earth). Observer Alice passes by 
with a velocity $v\ll c$. After few meters traveled (or, equivalently, 
after few seconds), Bob shouts ``Now!'' at Alice. Both Alice and Bob start
their stopwatches. Then, Alice suddenly stops walking away, makes a U-turn, and
goes sitting close  to Bob. They both shut their eyes and wait an
interval of time equal to $d/c$ before opening their eyes again. Since $v\ll c$,
their proper times are the same, 
their stopwatches are synchronized, their distance from Andromeda is the same 
($d$) and the time they have to wait before opening their eyes is also the
same ($d/c$). What will they 
see when they open their eyes? Bob will surely see events in Andromeda that 
were simultaneous with Bob's present when he yelled ``Now!'', and Alice will 
surely see events in Andromeda that were simultaneous with her present when 
Bob yelled ``Now!''. But, according to the Lorentz transformation of the time
coordinate, the Andromeda events that were simultaneous with Alice's present
when Bob yelled ``Now'' are $\simeq vd/c^2$ subsequent in time to those
simultaneous with Bob's present when he yelled ``Now!''.

This is a bizarre situation: people staying in the same place at the same
time and staring at the same source in the sky see different events. But,
this is a strict logical consequence of the accepted laws of
physics\footnote{Lorentz transformations intended as 
physical laws.}. Moreover, the very same logic could be
applied to the past, namely to events that happened millions of years ago.
Today, we should see a rainbow of different and simultaneous events while
observing the  Andromeda galaxy. Millions of years ago, we were not yet born and
it would be difficult to define the velocity $v$ of the observers then 
(and even just define the `observers'), but we are sure that the reader 
has got the point.

By the way, eq.~(\ref{eq2}) should have an 
observable consequence today: it is possible to demonstrate that even the 
periodic movement of the Earth around the Sun should induce a
sort of visible (wild and haphazard) `Doppler oscillations' of the radiation 
coming from very distant astronomical sources. Note that the frequency shift 
we are referring to here is not the standard Doppler shift due to the (usually 
high) relative velocity between the source and the observer. It is an
exclusively relativistic effect.
For the sake of derivation, let us focus on the velocity variation of the 
Earth with respect to the Andromeda galaxy during the Earth revolution around 
the Sun. The galaxy is assumed to be at rest with respect to the Sun. Let us 
further consider only a small trait of the Earth orbit, where the change of 
velocity (acceleration $a$) can be taken as constant (and obviously 
$a\Delta t \ll c$, for every $\Delta t$ considered). The acceleration is
assumed to be directed along the line of sight. Suppose that at initial 
Earth time instant $t_{E_1}$ the velocity of the Earth relative to 
Andromeda is equal to zero and then the simultaneous instant of time in 
Andromeda is $t_{A_1}=t_{E_1}$. At Earth time instant $t_{E_2}$, the relative 
velocity of the Earth has increased to $a(t_{E_2}-t_{E_1})=a\Delta t_E$ and 
thus, by making use of the differential form of eq.~(\ref{eq2}), we have the 
following result for the interval of time elapsed in Andromeda corresponding 
to the interval of time $\Delta t_E$ elapsed on the Earth

\begin{equation}
dt_A\simeq dt_E+\frac{d}{c^2}dv\quad\to\quad \Delta t_A\simeq \Delta t_E+
\frac{d}{c^2}a\Delta t_E=\Delta t_E\left(1+\frac{ad}{c^2}\right),
\label{eq4}
\end{equation}
where $d$ is, as before, the distance between the Earth and the Andromeda
galaxy.

Now, if there is a star in Andromeda that emits radiation at a frequency $\nu_0$
for an interval of time $\Delta t_A$, then it will emit a number of periods 
equal to $\nu_0\Delta t_A$. This number of periods will also be observed 
here on the Earth (after the traveling time $d/c$) but as emitted in the
shorter time interval $\Delta t_E$ and thus the frequency of the radiation
seen here on the Earth will be higher, equal to

\begin{equation}
\nu_E=\nu_0\left(1+\frac{ad}{c^2}\right).
\label{eq5}
\end{equation}

Since the motion of the Earth is not uniform around the Sun and since there 
are other dynamical mechanisms that contribute to the relative motion of the 
Earth with respect to distant galaxies (e.g.~motion of the Sun around the 
center of the Galaxy, relative motion of galaxies, not to mention the proper 
motion of the stars that emit radiation from inside the Andromeda galaxy), 
we should observe the light of distant galaxies weirdly and haphazardly 
Doppler shifted. The effects we would observe today would be due to radiation 
emitted a very long time ago ($\sim d/c$, where $d$ is the astronomical 
distance of the source from the Earth), but this delay does not cancel out
the phenomenon, we simply do not see it live.

\section{Muon decay and length contraction}
\label{sec:2}

In the '40s, studies conducted on muons generated by cosmic rays in the
upper atmosphere suggested that what was thought to be an anomalous 
absorption of these particles by the atmosphere itself was in fact due to 
their spontaneous decay and that the decay-rate depended upon muons' 
momentum~\cite{ros,fri}. The decay-rate dependence on momentum 
has been interpreted in the framework of special relativity as one of the 
neatest experimental verification of the time dilation of a `moving clock'.
Although muons mean lifetime $\tau_0$ is of only $\sim 2.2\,\mu s$ and 
thus not enough to guarantee their arrival at the lower atmosphere,
their high abundance at this atmospheric depth is explained by the fact that
their lifetime measured in the reference frame of the Earth is
relativistically dilated to $\tau=\tau_0/\sqrt{1-v^2/c^2}$ (due to their high 
relative velocity, $v\simeq 0.99\,c$); this is just the amount needed to 
explain their anomalous lower atmosphere abundance.

But, how is the same phenomenon explained when it is seen in the reference 
frame of the traveling muon? In the muon's rest frame, the particle 
decays, on average, after a time $\tau_0$, and from its perspective the rate 
of clocks on the Earth is slowed down. Therefore, from its perspective, 
an observer on the Earth should measure a {\em decrease} and not an 
{\em increase} of its lifetime and thus a decrease and not an increase in
the number of muons in the lower atmosphere. However, all relativists explain 
the phenomenon simply by invoking the length contraction of the atmosphere:
for a muon, the atmosphere is thinner and the particle has the time to
penetrate it deeper.

At first sight, this explanation appears quite neat and it is considered as
a solid proof of the internal coherence and strength of special relativity.
Under close inspection, however, it is a bit problematic and, apparently, it has
never been recognized as such before. 

\begin{figure}[t]
\begin{center}
  \includegraphics[width=8cm]{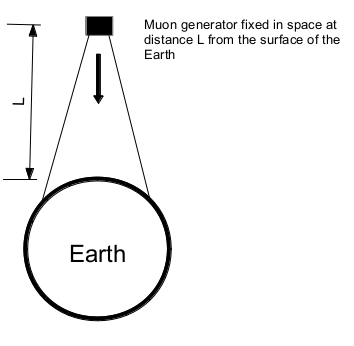}
\end{center}
\caption{Setup described in the text.}
\label{fig1}
\end{figure}

For the sake of simplicity, consider the setup shown in Figure~\ref{fig1}.
With regard to the key aspects of the process, it is completely equivalent
to the process observed in nature. 
The proper mean lifetime of a muon is $\tau_0$. This means 
that if we travel with the muon we will see it decay after an interval of time 
$\tau_0$. Observers on the Earth, instead, see the muon decay after a dilated 
interval of time $\tau= \tau_0/\sqrt{1-v^2/c^2}$, since $v\simeq c$. 
During this time, for the observers on the Earth, the muon travels a distance
$L=v\cdot\tau=v\cdot\tau_0/\sqrt{1-v^2/c^2}$. For the sake of argument,
the muon generator has been placed exactly at distance $L$ from the surface 
of the Earth and thus muons can reach the surface just before decay. 

In the reference frame of the muon, however, the particle sees the Earth 
approaching at speed $v$ and thus, from its point of view, during that 
interval of time the distance covered by the Earth before muon decay is 
$v\cdot\tau_0<L$. Namely, the muon disintegrates before touching the surface of 
the Earth. This result simply comes from elementary kinematics. Here, we 
only appeal to the principle of relativity by which the laws of physics 
(e.g.~kinematics) are the same in every inertial frame.

The only possibility to reconcile these two different views is the widely 
known and accepted explanation~(e.g.~see \cite{ss}) that in the
muon reference frame the distance that separates the muon (generator) from the
surface of the Earth  is Lorentz contracted, 

\begin{equation}
L'=L\cdot \sqrt{1-v^2/c^2}=v\cdot\tau_0/\sqrt{1-v^2/c^2}\cdot\sqrt{1-v^2/c^2}=
v\cdot\tau_0.
\label{eq3}
\end{equation}

\begin{figure}[t]
\begin{center}
  \includegraphics[width=9cm]{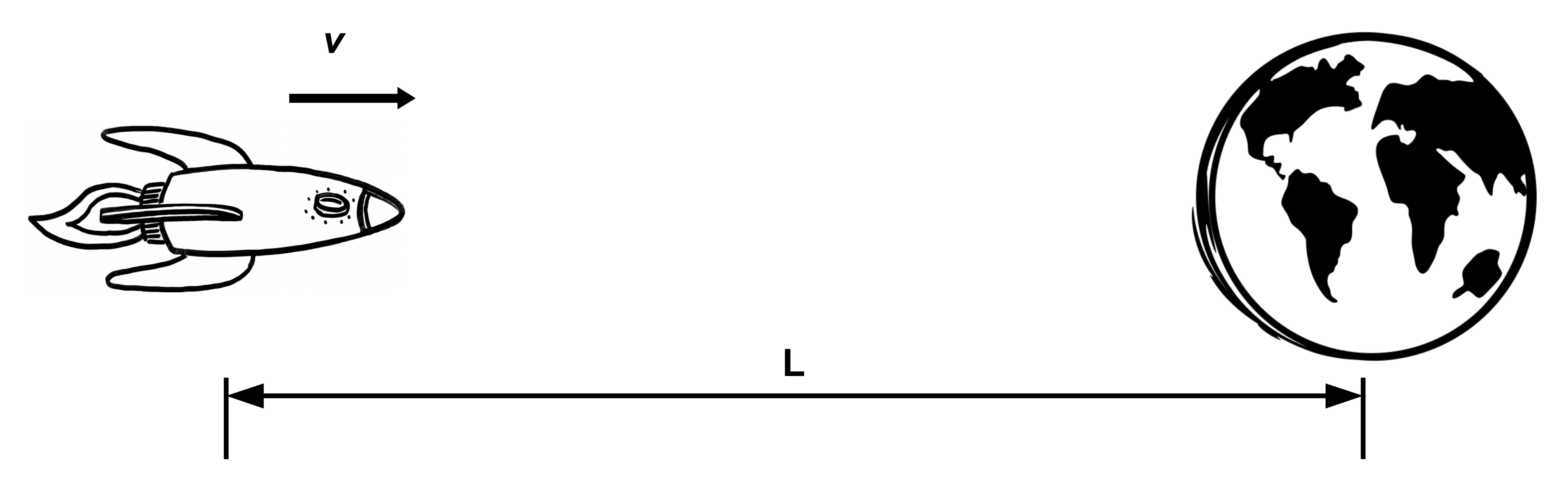}
\end{center}
\caption{Muons as interstellar travellers.}
\label{fig2}
\end{figure}

This means that, from muon's perspective, the Earth's surface appears to be
(and actually is) closer than $L$. This effect is also considered when
an interstellar journey of a spaceship traveling at speeds close to
that of light is analyzed from the perspective of the astronaut.
From the perspective of the
Earth, time on spaceship dilates and the astronaut can cover a huge distance
in a relatively short period of his own time. From the perspective of 
the astronaut, though, his time rate does not change and the only possibility 
to match the observations made by the observers on the Earth is that the 
distance to travel actually shortens for the astronaut.
Consider the situation depicted in Figure~\ref{fig2}. A spaceship is located
at a distance $L$ from the Earth and heads towards our planet at constant
velocity $v\lesssim c$. The distance $L$ is intended as measured from the Earth.
Suppose that the time $L/v$ needed by the spaceship to reach us is
greater than 100 years. According to special relativity, if $v$ is suitably
high, the observers on the Earth will measure a time dilation within the
spaceship that makes it possible for the astronaut to
reach the Earth in a shorter period of his own time, say 8 years.
Now, in the reference frame of the astronaut, the same outcome can only be
explained with length contraction. In order for the astronaut to reach
the Earth in 8 years of his own time, the distance $L$ should be suitably
shorter from his perspective. 
The other possibility, namely that the Earth appears faster to the astronaut,
cannot be accepted owing to the principle of relativity.
The very same principle of relativity, however, discloses a problem
with the length contraction explanation. According to this principle, there
is no reason to believe that the spaceship moves and the Earth is a rest. It
may well be the other way around. In that case, it should be the distance
seen by the Earth to be contracted. At any rate, the distance measured from the
Earth should be equal to that measured by the astronaut in the reference frame
of the spaceship because nobody can say who is moving and who is at rest.

Now, in the previous paragraph change the words `astronaut' or `spaceship'
with `muon' and it should be evident why length contraction cannot be an
acceptable explanation for the muon problem when it is analyzed in the muon
rest frame.

To recapitulate, two observations follow in order. First,
if the principle of relativity holds, one may equivalently assume that
the Earth is actually moving towards the muon and thus the distance $L$ that
separates us from the place where the particle originated (generator) is
already `shrunk'. Or, better, for the principle of relativity if we 
measure a distance equal to $L$, then also the muon must see the Earth 
distant $L$ from itself. Owing to the principle of relativity, the 
two distances (contracted or not) must be equal. {\em Length contraction,
like time dilation, is symmetrical\footnote{Unless we want 
to resort to Lorentz ether theory.} when the relative 
velocity is uniform, as is in this case}. 
Thus, the standard explanation of the muon lifetime dilation from 
muon's perspective becomes inconsistent, to say the least.

Secondly, what about observations of astronomical objects (matter) moving 
towards our position at relativistic speeds? Consider, for instance, 
relativistic jets of particles moving towards our position from Active 
Galactic Nuclei (AGNs)\footnote{Consider, for instance, the pulsar IGR 
J11014-6103: the estimated speed of its jet is $0.8c$.}.
According to the principle of relativity and the relativity of uniform
motion, we may equivalently consider the Solar System (or our galaxy) as
moving towards the jet particles at relativistic speeds and thus, according
to special relativity, these jets should appear to us a lot closer than 
the AGNs that have generated them. If the length contraction explanation of 
the muon decay phenomenon has a real physical meaning (it is physically real), 
then we should observe a weird distribution of matter in deep space, due to 
the existence of objects with different (and relativistic) relative velocity 
with respect to our reference frame. 

\section{Conclusion}
\label{sec:3}

We have shown that some primary special relativity effects, which are 
believed to be hardly detectable in everyday life, like time dilation, 
relativistic Doppler effect, and length contraction, should tangibly and
spectacularly show up here on the Earth: they should occur in ordinary
observations of known astronomical phenomena, also when the observations involve
astronomical systems that move with very low relative velocities ($v\ll c$) but
are placed at huge distances $d$ from us (with $d/c^2\gtrsim 1\,$s$^2$/m). 
In that regard, we have offered two examples: the first involves the so-called
Andromeda paradox, and the second, inter alia, calls into question the 
standard special relativity explanation of the muon lifetime dilation 
when the phenomenon is analyzed from muon's perspective. These two examples 
ultimately imply that if special relativity consequences (basically the
consequences deriving from the application of the Lorentz transformations) 
are real physical consequences, then the observable Universe should appear
very differently from what we actually see every clear night.
Unfortunately, none of the effects described in this paper, and that 
necessarily and strictly follows from special relativity, seem to have been 
ever observed. Thus, there are concrete elements to believe that 
something is actually not as it should be in the physical interpretation of 
Lorentz transformations and the allegedly real physical consequences of 
special relativity. A discussion on this last aspect from different standpoints 
can be found in~\cite{prob}. We want to end this paper with two quotes 
from the renowned physicist Mendel Sachs that appear to be particularly 
pertinent here: 

\begin{quote}
``I believe that Einstein's identification of the Lorentz transformation
with a {\em physical cause-effect} relation, and the subsequent conclusion 
about asymmetric ageing, {\em was a flaw}, not in the theory of relativity 
itself, [...], but rather {\em a flaw in the reasoning that Einstein 
used in this particular study}--leading him to an inconsistency with the 
meaning of space and time, according to his own 
theory.~\cite{sc}''~[emphasis added]
\end{quote}

\begin{quote}
``The crux of my argument was that the essence of Einstein's theory implies 
that the space-time transformations between relatively moving frames of 
reference must be interpreted strictly kinematically, rather than dynamically. 
Thus, according to this theory, the transformations are not more than the 
necessary scale changes that must be applied to the {\em measures} of space 
and time, when comparing the expressions of the laws of nature in relatively 
moving frames of reference, so as to satisfy the principle of relativity--that 
is, to ensure that their expressions in the different reference frames are in 
one-to-one correspondence.~\cite{sc2}''~[emphasis in the original text] 

\end{quote}

\section*{Acknowledgments}
The author acknowledges the anonymous reviewer for valuable comments and
suggestions.

\appendix

\section{Rigorous derivation of the equations~(\ref{eq4})
  and~(\ref{eq5})}

We take Einstein's derivation of the time dilation formula for a clock
moving in arbitrary motion (clock moving in a polygonal or continuously
curved line~\cite{e05}) and apply it to the case of a system moving on a
straight line but subject to a uniform acceleration $a$ for a short period of
time.
Hereafter, without loss of generality, we assume that all the involved
velocities are such that $v\ll c$. We also adopt the same assumption made by
Einstein in~\cite{e07} (and, implicitly, in~\cite{e05}), namely that
acceleration $a$ has negligible physical effects on the rate of clocks in the
accelerated frame. That is known as the `clock hypothesis'~\cite{clo}.

We shall see that when acceleration $a$ goes to zero, one recovers the
well-known Einstein's time dilation formula. On the other hand, if the distance
between the inertial observer and the accelerating system is suitably large,
one obtains the sought formulas.

Consider a moving reference frame $S'$ and an inertial (stationary) reference
frame $S$. Primed quantities refer to the system $S'$, while
non-primed ones refer to $S$. Moreover, $S'$ moves in the positive
$x$-direction of $S$, and all the three coordinate axes are parallel. Suppose
that $S'$ initially moves with constant velocity $v_1$, and at time $t=t'=0$,
the origins of $S$ and $S'$ overlap. Thus, the relation between the instants
of time $t_1'$ of $S'$ and $t_1$ of $S$ is given by the Lorentz trasnformation
of the time coordinate as follows

\begin{equation}
 t_1'=\frac{t_1 - \frac{v_1(v_1t_1)}{c^2}}{\sqrt{1-\frac{v_1^2}{c^2}}},
\label{eq6}
\end{equation}
since $x_1=v_1t_1$.

At instant $t_1$, the system $S'$ starts to accelerate in the positive or
negative $x$-direction with constant acceleration $a$, and at instant $t_2$
returns to uniform motion with the new constant velocity $v_2=v_1\pm
a(t_2-t_1)$.

Thus, the relation between the instants of time $t_2'$ of $S'$ and $t_2$ of
$S$ is now given by

\begin{equation}
  t_2'=\frac{t_2 - \frac{[v_1\pm a(t_2-t_1)][v_1t_1 + v_1(t_2-t_1) \pm
        \frac{1}{2}a(t_2-t_1)^2]}{c^2}}{\sqrt{1-\frac{[v_1\pm
          a(t_2-t_1)]^2}{c^2}}},
\label{eq7}
\end{equation}
where $x_2=v_1t_1 + v_1(t_2-t_1)\pm \frac{1}{2}a(t_2-t_1)^2$.

The interval of time $\Delta t'=t_2'-t_1'$ is thus equal to

\begin{equation}
  \Delta t'= \frac{t_2 - \frac{[v_1\pm a(t_2-t_1)][v_1t_1 + v_1(t_2-t_1)\pm
        \frac{1}{2}a(t_2-t_1)^2]}{c^2}}{\sqrt{1-\frac{[v_1\pm
          a(t_2-t_1)]^2}{c^2}}} - \frac{t_1 -
    \frac{v_1(v_1t_1)}{c^2}}{\sqrt{1-\frac{v_1^2}{c^2}}}.
\label{eq8}
\end{equation}

Now, it is not difficult to see that if  we set $a=0$ in equation (\ref{eq8})
and do not neglect terms containing the 2nd power of $v/c$, we recover
Einstein's time dilation formula

\begin{equation}
  \Delta t'= \Delta t \sqrt{1-\frac{v_1^2}{c^2}}.
  \label{eq9}
\end{equation}

On the other hand, if we set $v_1t_1=d$, with $d$ equal to an extremely large
astronomical distance, and if we consequently adopt the natural
approximations, $v_1(t_2-t_1)\pm \frac{1}{2}a(t_2-t_1)^2\ll v_1t_1$ and 
$\frac{[v_1\pm a(t_2-t_1)]^2}{c^2}\approx\frac{v_1^2}{c^2}\approx 0$ (we are
now neglecting again terms containing the 2nd power of $v/c$),
from equation~(\ref{eq8}) we arrive at the following relation

\begin{equation}
  \Delta t'= \Delta t \left[1-\frac{\pm ad}{c^2} \right],
\label{eq10}
\end{equation}
which is the sought formula for the time dilation/contraction (generalization
of equation~(\ref{eq4}) in the text).

In short, we have replicated Einstein's derivation of the time dilation for a
clock arbitrarily moving with respect to a stationary clock~\cite{e05}. Like
Einstein, we started from the Lorentz transformation of the time coordinate.
However, we have plugged in the equation an explicit and simpler type of motion
for the moving clock: namely, the moving clock moves away from the stationary
one on a straight line at constant velocity $v_1$ for a time $t_1$, and then,
for a time $(t_2-t_1)$, it accelerates with a low acceleration $a$. That is
simpler than Einstein's motion in a polygonal or continuously curved
line~\cite{e05}. Therefore, if special relativity holds for non-uniform motion
in a ``continuously curved line'', it does hold also for a body slightly
accelerating in a straight line. By the way, what we have done so far is
equivalent to mapping the considered set-up onto a continuous sequence of
events which are analyzed with respect to instantaneous co-moving inertial
frames.

Now, suppose that during interval $\Delta t'$, the light source at rest in
$S'$ emits a beam of light of frequency $\nu'$. That means that $N$ wave
crests are emitted with $N=\nu'\Delta t'$. The same number of crests
must then be received by the observer in $S$ exactly after the traveling
time $d/c$, no matter how big $d/c$ is.
Moreover, the observer in $S$ will receive the $N$ wave crests within the
shorter interval of time $\Delta t$ because, for $S$, the whole emission
process in $S'$ has taken place within $\Delta t$ (the traveling time $d/c$
cannot affect that duration since $d/c$ is only a delay in receiving the
wave train).
That means that the observer in $S$ receives a beam of light of frequency $\nu$
such that $\nu\Delta t = N =\nu'\Delta t'$, and thus 

\begin{equation}
  \nu=\nu'\left[1-\frac{(\pm a)d}{c^2}\right],
\label{eq11}
\end{equation}
which is the generalization of equation~(\ref{eq5}) in the text.

Let us remind that the above formula gives a frequency shift that has nothing
to do with the standard Doppler shift depending upon relative speed nor with
the gravitational redshift depending upon gravity.

\end{document}